# Parallel mathematical models of dynamic objects

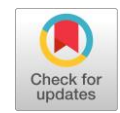

Roman Voliansky [a,1,*], Andri Pranolo [b,2]

[a] Electric Engineering Department, Dniprovsky State Technical University, Kamyanske, Ukraine  
[b] Department of Informatics, Universitas Ahmad Dahlan, Yogyakarta, Indonesia  
[1] voliansky@ua.fm; [2] andri.pranolo@tif.uad.ac.id  
* corresponding author

ABSTRACT

The paper deals with the developing of the methodological backgrounds for the modeling and simulation of complex dynamical objects. Such backgrounds allow us to perform coordinate transformation and formulate the algorithm of its usage for transforming the serial mathematical model into parallel ones. This algorithm is based on partial fraction decomposition of the transfer function of a dynamic object. Usage of proposed algorithms is one of the ways to decrease calculation time and improve PC usage while a simulation is being performed. We prove our approach by considering the example of modeling and simulating of fourth order dynamical object with various eigenvalues. This example shows that developed parallel model is stable, well-convergent, and high-accuracy model. There is no defined any calculation errors between well-known serial model and proposed parallel one. Nevertheless, the proposed approach's usage allows us to reduce calculation time by more than 20% by using several CPU's cores while calculations are being performed.

## 1. Introduction

Nowadays different mathematical models are widely used in various scientific, economic, social, industrial and other applications [1]–[3]. Various kinds of mathematical models make it possible such as to predict some events and actions [4] or make conclusions [5], and process some data and find interrelations between its various pieces [6]–[8].

One of the most important spheres of mathematical modeling and simulation is its scientific application [9]–[13]. One can find that the benefits of mathematical models are very useful while novel objects, properties, and characteristics are being created [14]–[16] and/or it is impossible to perform some measurements. Moreover, tons of mathematical models allow studying objects which have not been created yet or other ways of their study are very dangerous and can cause accidents [17] immediately or after a while.

Class of dynamic models is a very wide class of mathematical models which allow us to study object's actions in process of time. Usage of such models makes it possible to study various dynamic and static characteristics and predict a behavior of the studied object. If one performs analysis of known dynamical models, he can find that these models are parallel if a considered object has parallel channels for transmitting energy from inputs to outputs, and these models are series otherwise. Many models of robotic devices [18] and some complex industrial processes are parallel models due to using many inputs and outputs but, nevertheless, most of the dynamic models are series ones. One of the most powerful tools for studying of object's dynamic is Simulink from MatLab software [19]. This software is intended to solve various linear and nonlinear algebraic and differential equations which are given by state space equations as well as transfer functions. One can conclude that numerical solutions of differential equations are performed in a serial way. This fact determines the main drawback of differential equations' solvers which are integrated into Simulink. These solvers define solution of differential equations one after the other and use for this action only one core from multi-core CPU. This drawback increases calculation time dramatically and causes under-utilization of modern PC while a dynamic of the object is being studied. One can find similar problems with other mathematical software as well.

We offer to decrease calculation time and improve CPU usage by developing a scientific background for developing the parallel model that can be simulated by using different cores of CPU, CPUs, and PCs. Parallel computations are widely used while scientific research is being carried out [20], [21] but no parallel dynamic models were found while scientific publications were analyzed. Well-know Parallel Computational Toolbox which is part of Matlab does not allow to decompose model into small parallel pieces and calculate them in a parallel way.

## 2. Method

### 2.1. Transformation in Continuous Time Domain

We consider dynamic of a generalized dynamical object given by the following linear state space equations in matrix form [22].

$$\dot{\mathbf{X}} = \mathbf{AX} + \mathbf{BU}, \quad \mathbf{Y} = \mathbf{CX} + \mathbf{DU}, \tag{1}$$

where $\mathbf{X}$ is the $n \times 1$-sized state space vector, $\mathbf{Y}$ is the $m \times 1$-sized output vector, $\mathbf{U}$ is the $r \times 1$-sized vector of input efforts, $\mathbf{A}$ is the $n \times n$-sized state matrix, $\mathbf{B}$ is the $n \times r$-sized input matrix, $\mathbf{C}$ is the $m \times n$-sized output matrix and $\mathbf{D}$ is the $m \times r$-sized feed through matrix.

It is easy to define matrix transfer function by using (1)

$$\mathbf{W}(s) = \frac{\mathbf{Y}(s)}{\mathbf{U}(s)} = \mathbf{C}(s\mathbf{E} - \mathbf{A})^{-1}\mathbf{B} + \mathbf{D}, \tag{2}$$

Here, $\mathbf{E}$ is an identity matrix.

If one takes into account the matrix inversion rule, (2) can be rewritten as follows

$$\mathbf{W}(s) = \frac{\mathbf{Y}(s)}{\mathbf{U}(s)} = \mathbf{C}\frac{adj(s\mathbf{E} - \mathbf{A})}{det(s\mathbf{E} - \mathbf{A})}\mathbf{B} + \mathbf{D}, \tag{3}$$

where $det(s\mathbf{E} - \mathbf{A})$ is the characteristic polynomial of matrix $\mathbf{A}$, $adj(s\mathbf{E} - \mathbf{A})$ is the adjugate matrix to matrix $s\mathbf{E} - \mathbf{A}$

$$adj(s\mathbf{E} - \mathbf{A}) = \mathbf{F}^T, \quad \mathbf{F}_{ij} = (-1)^{i+j}\mathbf{M}_{ij}, \tag{4}$$

here $\mathbf{M}_{ij}$ is the (i,j) minor of $s\mathbf{E} - \mathbf{A}$ matrix.

Finally, we reduce (3) to a common denominator and write down following matrix transfer function

$$\mathbf{W}(s) = \frac{\mathbf{Y}(s)}{\mathbf{U}(s)} = \frac{\mathbf{C} \cdot adj(s\mathbf{E} - \mathbf{A}) \cdot \mathbf{B} + det(s\mathbf{E} - \mathbf{A}) \cdot \mathbf{D}}{det(s\mathbf{E} - \mathbf{A})}. \tag{5}$$

Analysis of (5) makes it possible to formulate the following statements:

**Statement 1.** Generalized matrix transfer function of a considered linear object consists of two parts. The first one defines the controlled dynamic of the considered object and the second one defines the influence of control signal on observed state variables.

**Statement 2.** If state space equations have D matrix with non-zero elements, the order of numerator's and denominator's polynomials equal to object order.

We assume that A, B, C, D matrices are well-defined and a result of numerator's (5) calculation can be rewritten by using matrix polynomial

$$\mathbf{C} \cdot adj(s\mathbf{E} - \mathbf{A}) \cdot \mathbf{B} + det(s\mathbf{E} - \mathbf{A}) \cdot \mathbf{D} = \sum_{k=1}^{n} \boldsymbol{\alpha}_k s^k, \tag{6}$$

where $\boldsymbol{\alpha}_i$ is the some $m \times r$-sized matrix.

The denominator of (5) can be given in a similar way

$$det(s\mathbf{E} - \mathbf{A}) = \sum_{i=1}^{n} \beta_i s^i. \tag{7}$$

Expressions (6) and (7) give us possibility to rewrite (5)

$$\mathbf{W}(s) = \frac{\mathbf{Y}(s)}{\mathbf{U}(s)} = \frac{\sum_{k=1}^{n} \boldsymbol{\alpha}_k s^k}{\sum_{i=1}^{n} \beta_i s^i}. \tag{8}$$

Transfer function (8) makes it possible to write down following differential equation

$$\sum_{i=1}^{n} \beta_i s^i \cdot \mathbf{Y}(s) = \mathbf{U}(s) \cdot \sum_{k=1}^{n} \boldsymbol{\alpha}_k s^k, \tag{9}$$

which we call as matrix differential equation in canonical form.

Now we define eigenvalues of matrix **A** as a solution of the equation

$$det(\lambda \mathbf{E} - \mathbf{A}) = 0. \tag{10}$$

It is quite clear that equation (10) can have zero, real and complex eigenvalues. Moreover, these eigenvalues can be k-th multiple eigenvalues.

Now we assume that (10) has k0 multiple zero eigenvalues, k11 real eigenvalues of order k12 multiplicity and k21 complex conjugate eigenvalues of order k12 multiplicity. This assumption allows us to rewrite denominator (8) as follows

$$det(\lambda \mathbf{E} - \mathbf{A}) = \lambda^{k0} \prod_{i=1}^{k11} (\lambda - \lambda_i)^{k12} \prod_{j=1}^{k21} [(\lambda - \lambda_{j\,Re} - I\lambda_{j\,Im})(\lambda - \lambda_{j\,Re} + I\lambda_{j\,Im})]^{k22}, \tag{11}$$

where $\lambda_i$ is the $i$-th real eigenvalue, $\lambda_{j\,Re}$ and $\lambda_{j\,Im}$ are the real and imaginary parts of $j$-th complex conjugate eigenvalue, $I$ is the imaginary unit.

Expression (11) allows us to rewrite (8) as follows

$$\mathbf{W}(s) = \frac{\mathbf{Y}(s)}{\mathbf{U}(s)} = \sum_{k=1}^{n} \boldsymbol{\alpha}_k s^k \cdot \frac{1}{s^{k0} \prod_{i=1}^{k11}(s-\lambda_i)^{k12} \prod_{j=1}^{k21}\left[(s-\lambda_{jRe}-I\lambda_{jIm})(s-\lambda_{jRe}+I\lambda_{jIm})\right]^{k22}} \quad (12)$$

and decompose second multiplier in following way [23]

$$\frac{1}{\sum_{i=1}^{n} \beta_i s^i} = \sum_{i=1}^{k0} \frac{e_i}{s^i} + \sum_{i=1}^{k11}\sum_{j=1}^{k12} \frac{f_{ij}}{(s-\lambda^i)^j} + \sum_{i=1}^{k21}\sum_{j=1}^{k22} \frac{g_{ij}}{\left[(s-\lambda_{jRe}-I\lambda_{jIm})(s-\lambda_{jRe}+I\lambda_{jIm})\right]^j}, \quad (13)$$

where $e_i, f_{ij}, g_{ij}$ are some coefficients.

One can find these coefficients by solving the following equation

$$1 = \sum_{i=1}^{k0} e_i s^{k0-i} \prod_{i=1}^{k11}(s-\lambda_i)^{k12} \prod_{j=1}^{k21}\left[(s-\lambda_{jRe}-I\lambda_{jIm})(s-\lambda_{jRe}+I\lambda_{jIm})\right]^{k22}$$
$$+ \sum_{i=1}^{k11}\sum_{j=1}^{k12} \frac{f_{ij} s^{k0} \prod_{j1=1}^{k21}\left[(s-\lambda_{j1Re}-I\lambda_{j1Im})(s-\lambda_{jRe}+I\lambda_{jIm})\right]^{k22} \prod_{i1=1}^{i}\prod_{i1=i+1}^{k11}(s-\lambda_{i1})^{k12-j}}{(s-\lambda^i)^j} + \quad (14)$$
$$+ \sum_{i=1}^{k21}\sum_{j=1}^{k22} \frac{g_{ij} s^{k0} \prod_{i=1}^{k11}(s-\lambda_i)^{k12} \prod_{i1=1}^{i}\prod_{i1=i+1}^{k11}\left[(s-\lambda_{j1Re}-I\lambda_{j1Im})(s-\lambda_{jRe}+I\lambda_{jIm})\right]^{k22-j}}{\left[(s-\lambda_{jRe}-I\lambda_{jIm})(s-\lambda_{jRe}+I\lambda_{jIm})\right]^j}.$$

We offer to use (13) for rewriting (12)

$$\mathbf{W}(s) = \frac{\mathbf{Y}(s)}{\mathbf{U}(s)} = \sum_{k=1}^{n} \boldsymbol{\alpha}_k s^k \sum_{i=1}^{k0} \frac{e_i}{s^i} + \sum_{k=1}^{n} \boldsymbol{\alpha}_k s^k \sum_{i=1}^{k11}\sum_{j=1}^{k12} \frac{f_{ij}}{(s-\lambda^i)^j} +$$
$$+ \sum_{k=1}^{n} \boldsymbol{\alpha}_k s^k \sum_{i=1}^{k21}\sum_{j=1}^{k22} \frac{g_{ij}}{\left[(s-\lambda_{jRe}-I\lambda_{jIm})(s-\lambda_{jRe}+I\lambda_{jIm})\right]^j}. \quad (15)$$

We call (15) as parallel matrix transfer function.

Now we assume the existence of some virtual state variable $\mathbf{Yi}$ and define state vector $\mathbf{Y}$ in the following way

$$\mathbf{Y}(s) = \mathbf{Y1}(s) + \mathbf{Y2}(s) + \mathbf{Y3}(s). \quad (16)$$

This assumption makes it possible to define (15) as the sum of subsystem transfer functions

$$\mathbf{W}(s) = \frac{\mathbf{Y}(s)}{\mathbf{U}(s)} = \mathbf{W1}(s) + \mathbf{W2}(s) + \mathbf{W3}(s), \quad (17)$$

where

$$\mathbf{W1}(s) = \frac{\mathbf{Y1}(s)}{\mathbf{U}(s)} = \sum_{i=1}^{k0} e_i \sum_{k=1}^{n} \boldsymbol{\alpha}_k s^{k-i};$$

$$\mathbf{W2}(s) = \frac{\mathbf{Y2}(s)}{\mathbf{U}(s)} = \sum_{i=1}^{k11} \sum_{j=1}^{k12} \frac{f_{ij} \sum_{k=1}^{n} \boldsymbol{\alpha}_k s^k}{(s - \lambda^i)^j};$$

$$\mathbf{W3}(s) = \frac{\mathbf{Y3}(s)}{\mathbf{U}(s)} = \sum_{i=1}^{k21} \sum_{j=1}^{k22} \frac{g_{ij} \sum_{k=1}^{n} \boldsymbol{\alpha}_k s^k}{[(s - \lambda_{jRe} - I\lambda_{jIm})(s - \lambda_{jRe} + I\lambda_{jIm})]^j}.$$

(18)

It is clearly understood that the order of each summand (17) defines its order

$$n = k0 + k11 k12 + k21 k22.$$  (19)

Expressions (15) - (19) allows us to make the following statements:

**Statement 3.** Usage of partial fraction decomposition allows us to rewrite object's transfer function into parallel way by defining virtual parallel channels. Order each of them less than the order of whole transfer function.

**Statement 4.** One can study the dynamic of the generalized linear object in an easy way by studying the dynamic of all parallel channels.

We suggest determination of state space equations for a virtual parallelized object by using (16)-(18) as follows

$$\mathbf{Y1}(s) = \sum_{i=1}^{k0} e_i \sum_{k=1}^{n} \boldsymbol{\alpha}_k s^{k-i} \mathbf{U}(s);$$

$$\sum_{i=1}^{k11} \sum_{j=1}^{k12} \mathbf{Y2}(s)(s - \lambda^i)^j = \sum_{i=1}^{k11} \sum_{j=1}^{k12} f_{ij} \sum_{k=1}^{n} \boldsymbol{\alpha}_k s^k \mathbf{U}(s);$$

$$\sum_{i=1}^{k21} \sum_{j=1}^{k22} \mathbf{Y3}(s)[(s - \lambda_{jRe} - I\lambda_{jIm})(s - \lambda_{jRe} + I\lambda_{jIm})]^j = \sum_{i=1}^{k21} \sum_{j=1}^{k22} g_{ij} \sum_{k=1}^{n} \boldsymbol{\alpha}_k s^k \mathbf{U}(s);$$

$$\mathbf{Y}(s) = \mathbf{Y1}(s) + \mathbf{Y2}(s) + \mathbf{Y3}(s).$$

(20)

We call (30) as parallel state space equations.

It is clearly understood that (30) contains $n$ independent differential equations which can be solved separately analytically or numerically. If one solves them in a numerical way he can use different cores and CPUs or even different PCs to solve (30). When values of virtual state space vectors $\mathbf{Yi}$ are obtained, one can store and sum them for getting values of $\mathbf{Y}$ vector. This is the main benefit of the proposed approach which can be implemented by means of computer technology in an easy way.

## 2.2. State space equations in the discrete time domain

There are tons of various numerical methods for solving ordinary differential equations which differs with various approximations of the derivative operator. The common feature of these methods is a usage of state variables in previous, current, and future time.

Thus, we consider the generalized approximation of derivative operator

$$s = s(z^{-q1}, z^{-q1+1}, \ldots, z^{-1}, 1, z, \ldots, z^{q2-1}, z^{q2}),$$  (21)

where $z$ is a shift operator, $q1$ is a number of previous time's samples, $q2$ is a number of future time's samples.

Usage of (21) allows us to rewrite (20) in discrete time domain

$$\mathbf{Y1}(z) = \sum_{i=1}^{k0} e_i \sum_{k=1}^{n} \mathbf{\alpha}_k s(z^{-q1}, z^{-q1+1}, \ldots, z^{-1}, 1, z, \ldots, z^{q2-1}, z^{q2})^{k-i} \mathbf{U}(z);$$

$$\sum_{i=1}^{k11} \sum_{j=1}^{k12} \mathbf{Y2}(z) \left( s(z^{-q1}, z^{-q1+1}, \ldots, z^{-1}, 1, z, \ldots, z^{q2-1}, z^{q2}) - \lambda^i \right)^j =$$

$$= \sum_{i=1}^{k11} \sum_{j=1}^{k12} f_{ij} \sum_{k=1}^{n} \mathbf{\alpha}_k s(z^{-q1}, z^{-q1+1}, \ldots, z^{-1}, 1, z, \ldots, z^{q2-1}, z^{q2})^k \mathbf{U}(z);$$

$$\sum_{i=1}^{k21} \sum_{j=1}^{k22} \mathbf{Y3}(z) \left[ \left( s(z^{-q1}, z^{-q1+1}, \ldots, z^{-1}, 1, z, \ldots, z^{q2-1}, z^{q2}) - \lambda_{jRe} - I\lambda_{jIm} \right) \times \right.$$

$$\left. \times \left( s(z^{-q1}, z^{-q1+1}, \ldots, z^{-1}, 1, z, \ldots, z^{q2-1}, z^{q2}) - \lambda_{jRe} + I\lambda_{jIm} \right) \right]^j =$$

$$= \sum_{i=1}^{k21} \sum_{j=1}^{k22} g_{ij} \sum_{k=1}^{n} \mathbf{\alpha}_k s(z^{-q1}, z^{-q1+1}, \ldots, z^{-1}, 1, z, \ldots, z^{q2-1}, z^{q2})^k \mathbf{U}(z);$$

$$\mathbf{Y}(z) = \mathbf{Y1}(z) + \mathbf{Y2}(z) + \mathbf{Y3}(z).$$

(22)

Equations (22) allow us to formulate the following statement.

**Statement 5.** Solver, which calculates dynamic for every parallel channel, operates with pre and post information about virtual state variables' values and control efforts in the general case. One can use (22) for defining virtual and real state variables for various sample time $T$.

## 3. Results and Discussion

### 3.1. Modeling of DC Electric Drive in Parallel Way

#### 3.1.1. Parallel model of DC electric drive in continuous time

It is well-known fact that one can describe dynamic of DC electric drive by the following differential equations which are given by using the relative units

$$sx_1 = a_{12}x_2; \quad sx_2 = a_{23}x_3; \quad sx_3 = a_{32}x_2 + a_{33}x_3 + a_{34}x_4;$$
$$sx_4 = a_{44}x_4 + b_4 U; \quad y = x_1$$

(23)

where coefficients $a_{ij}$ are defined as follows

$$a_{12} = 1; \; a_{23} = \frac{1}{T_m}; \; a_{32} = a_{33} = -\frac{1}{T_e}; \; a_{34} = \frac{1}{T_e};$$
$$a_{44} = -\frac{1}{T_c}; \; b_4 = -a_{44}; \quad T_m = \frac{JR}{c^2}; \; T_e = \frac{L}{R},$$

(24)

here $J$ is a rotor inertia, $R$ is an armature resistance, $c$ is a back-emf constant, $L$ is an armature inductance, $T_c$ is power converter coefficient.

We rewrite (23) into matrix form

$$s\mathbf{X} = \mathbf{AX} + \mathbf{BU}, \; y = x_1,$$

(25)

where

$$X = (x_1 \ x_2 \ x_3 \ x_4)^T; U = (0 \ 0 \ 0 \ U)^T;$$
$$B = (0 \ 0 \ 0 \ b_4); A = \begin{pmatrix} 0 & a_{12} & 0 & 0 \\ 0 & 0 & a_{23} & 0 \\ 0 & a_{32} & a_{33} & a_{34} \\ 0 & 0 & 0 & a_{44} \end{pmatrix}$$
(26)

and define transfer function in the following way

$$W(s) = \frac{y(s)}{U(s)} = \frac{a_{12}a_{23}a_{34}b_4}{s^4 - (a_{33}+a_{44})s^3 + (a_{33}a_{44} - a_{23}a_{32})s^2 + a_{44}a_{32}a_{23}s}.$$
(27)

Let us consider characteristic polynomial

$$D(\lambda) = \lambda^4 - (a_{33}+a_{44})\lambda^3 + (a_{33}a_{44} - a_{23}a_{32})\lambda^2 + a_{44}a_{32}a_{23}\lambda$$
(28)

and define its eigenvalues

$$\lambda_1 = 0; \lambda_2 = a_{44}; \lambda_{3,4} = 0.5a_{33} \pm 0.5\sqrt{4a_{23}a_{32} + a_{33}^2}.$$
(29)

Eigenvalues $\lambda_3, \lambda_4$ are complex conjugated ones if the following condition is true

$$4a_{23}a_{32} + a_{33}^2 < 0.$$
(30)

One can find that this condition is true for the most industrial application, so we assume that (28) has two complex conjugate eigenvalues

$$\lambda_{3,4} = 0.5a_{33} \pm 0.5I\sqrt{|4a_{23}a_{32} + a_{33}^2|}.$$
(31)

This fact makes it possible to perform for (28) partial fraction decomposition

$$\frac{a_{12}a_{22}a_{34}b_4}{s^4 - (a_{33}+a_{44})s^3 + (a_{33}a_{44} - a_{23}a_{32})s^2 + a_{44}a_{32}a_{23}s} = \frac{k0}{s} + \frac{k1}{s-a_{44}} +$$
$$+ \frac{k2s + k3}{\left(0.5a_{33} + 0.5I\sqrt{|4a_{23}a_{32} + a_{33}^2|}\right)\left(0.5a_{33} - 0.5I\sqrt{|4a_{23}a_{32} + a_{33}^2|}\right)}$$
(32)

or

$$\frac{a_{12}a_{22}a_{34}b_4}{s^4 - (a_{33}+a_{44})s^3 + (a_{33}a_{44} - a_{23}a_{32})s^2 + a_{44}a_{32}a_{23}s} = \frac{k0}{s} + \frac{k1}{s-a_{44}} + \frac{k2s+k3}{s^2 - a_{33}s - a_{23}a_{32}}.$$
(33)

We offer to perform for (33) some algebraic transformations and simplifications which allows us to write down equations for defining unknown $ki$ coefficients

$$k0 + k1 + k2 = 0; \quad a_{33}(k0+k1) + a_{44}(k0+k2) - k3 = 0;$$
$$a_{23}a_{32}(k0+k1) + a_{44}(k3 - k0a_{33}) = 0; \quad a_{32}a_{44}k_0 = a_{12}a_{34}b_4$$
(34)

and define them in the following way

$$k0 = \frac{a_{12}a_{34}b_4}{a_{32}a_{44}}; \quad k_1 = -\frac{a_{12}a_{23}a_{34}b_4}{a_{44}(a_{23}a_{32} + a_{33}a_{44} - a_{44}^2)};$$
$$k_2 = -\frac{(a_{33} - a_{44})a_{12}a_{34}b_4}{a_{32}(a_{23}a_{32} + a_{33}a_{44} - a_{44}^2)}; \quad k_3 = \frac{a_{12}a_{34}b_4(a_{23}a_{32} + a_{33}^2 - a_{33}a_{44})}{a_{32}(a_{23}a_{32} + a_{33}a_{44} - a_{44}^2)}.$$
(35)

We use (33) to write down parallel state space equations as follows

$$sy1 = k0U;$$
$$sy2 = a_{44}y2 + k1U;$$
$$s^2 y3 - a_{33}sy3 - a_{23}a_{32}y3 = k2sU + k3U$$
(36)

or

$$sy1 = k0U; \quad sy2 = a_{44}y2 + k1U;$$
$$sy31 = y32; \quad sy32 = a_{33}y32 + a_{23}a_{32}y31 + k2sU + k3U;$$
$$y = y1 + y2 + y31.$$
(37)

It is clearly understood that the transformed mathematical model has three parallel channels. Two of them have first order and the third one is a dynamical object of second order. In such a way we replace one complex fourth order dynamical object with three simple dynamical objects.

### 3.1.2. Parallel model of DC electric drive in the discrete time domain

Now we transform (37) into the discrete time domain by applying bilinear transformation which allows us to approximate derivative operator in the following way

$$s \approx 2(1 - z^{-1})/(1 + z^{-1})/T$$
(38)

and rewrite (37)

$$\frac{2}{T}\frac{1-z^{-1}}{1+z^{-1}}y1 = k0U; \quad \frac{2}{T}\frac{1-z^{-1}}{1+z^{-1}}y2 = a_{44}y2 + k1U;$$
$$\frac{2}{T}\frac{1-z^{-1}}{1+z^{-1}}y31 = y32; \quad \frac{2}{T}\frac{1-z^{-1}}{1+z^{-1}}y32 = a_{33}y32 + a_{23}a_{32}y31 + k2\frac{2}{T}\frac{1-z^{-1}}{1+z^{-1}}U + k3U$$
(39)

or

$$y1 = z^{-1}y1 + \frac{Tk0}{2}(U + z^{-1}U); \quad y2 = \frac{1}{2 - a_{44}T}\left[(2 + a_{44}T)z^{-1}y2 + Tk1(U + z^{-1}U)\right];$$
$$y31 = z^{-1}y31 + \frac{T}{2}(y32 + z^{-1}y32);$$
$$y32 = \frac{\left[(2 + a_{33}T)z^{-1}y32 + Ta_{23}a_{32}(y31 + z^{-1}y31) + k2(U - z^{-1}U) + Tk3(U + z^{-1}U)\right]}{2 - a_{33}T}.$$
(40)

One can use (27) or similar approximations of the derivative operator for obtaining the mathematical model of DC electric drive in the discrete time domain. Right-hand expressions of (40) always depend on previous values of virtual state variables, but this dependence is different for various approximations.

It is well-known fact that these approximations define stability and accuracy of numerical solution (37). From a programming viewpoint, it is easy to solve so-called mesh equations.

### 3.1.3. Parallel mesh model of DC electric drive

Now we rewrite (40) as mesh equations

$$\begin{aligned}
y1[i] &= y1[i-1] + m_{1i}U[i] + m_{1(i-1)}U[i-1]; \\
y2[i] &= a_{22(i-1)}y2[i-1] + m_{2i}U[i] + m_{2(i-1)}U[i-1]; \\
y31[i] &= y31[i-1] + a_{32i}y32[i] + a_{32(i-1)}y32[i-1]; \\
y32[i] &= a_{42(i-1)}y32[i-1] + a_{41i}y31[i] + a_{41(i-1)}y31[i-1] + m_{4i}U[i] + m_{4(i-1)}U[i-1].
\end{aligned} \tag{41}$$

where

$$m_{1i} = \frac{Tk0}{2}; m_{1(i-1)} = \frac{Tk0}{2}; a_{22(i-1)} = \frac{2 + a_{44}T}{2 - a_{44}T}; m_{2i} = \frac{Tk1}{2 - a_{44}T}; m_{2(i-1)} = \frac{Tk1}{2 - a_{44}T};$$

$$a_{32i} = \frac{T}{2}; a_{32(i-1)} = \frac{T}{2}; \tag{42}$$

$$a_{41(i-1)} = \frac{2 + a_{33}T}{2 - a_{33}T}; a_{41i} = \frac{Ta_{23}a_{32}}{2 - a_{33}T}; a_{42i} = \frac{Ta_{23}a_{32}}{2 - a_{33}T}; m_{4i} = \frac{k2 + Tk3}{2 - a_{33}T}; m_{4(i-1)} = \frac{Tk3 - k2}{2 - a_{33}T}$$

and represent them as the matrix equation

$$\mathbf{A_d Y_d + M_d U_d = 0}, \tag{43}$$

here

$$\mathbf{M_d} = \begin{pmatrix} m_{1i} & m_{2i} & 0 & m_{4i} \\ m_{1(i-1)} & m_{2(i-1)} & 0 & m_{4(i-1)} \end{pmatrix}^T;$$

$$\mathbf{A_d} = \begin{pmatrix} -1 & 1 & 0 & 0 & 0 & 0 & 0 & 0 \\ 0 & 0 & -1 & a_{22(i-1)} & 0 & 0 & 0 & 0 \\ 0 & 0 & 0 & 0 & -1 & 1 & a_{32i} & a_{32(i-1)} \\ 0 & 0 & 0 & 0 & a_{41i} & a_{41(i-1)} & -1 & a_{42(i-1)} \end{pmatrix} \tag{44}$$

$$\mathbf{Y_d} = (y1[i] \quad y1[i-1] \quad y2[i] \quad y2[i-1] \quad y31[i] \quad y31[i-1] \quad y32[i] \quad y32[i-1])^T;$$

$$\mathbf{U_d} = (U[i] \quad U[i-1])^T;$$

Solution (43) for every sample time allows us to define virtual state variables $yi$ in the discrete time domain and then define output state variable $y$ as well.

$$y[i] = y1[i] + y2[i] + y31[i]. \tag{45}$$

One can use (43) and (45) for defining output variable of considered DC electric drive.

### 3.2. Simulation's Results

Results of numerical solutions (23) and (37) are shown in the Fig. 1.

We obtain these results in the discrete time domain with sample time $T = 10^{-5}$ s and parameters $a_{12} = 1\,s^{-1}; a_{23} = 50\,s^{-1}; a_{32} = a_{33} = -a_{34} = -125\,s^{-1}; a_{44} = -b_4 = -1000\,s^{-1}$.

Simulation time for the serial model is 0.4299s while one CPU core is being used. First equation of the parallel model is solved in 0.0878s, the second equation is solved in 0.1347s, and the last two equations are solved in 0.2615s. It needs 0.0651s for summing the results of each parallel channel. So, simulation time for the parallel model is 0.3266s by using 3 CPU cores. Thus, we decrease simulation time by 24% while we use parallel model and its multi-core program implementation.

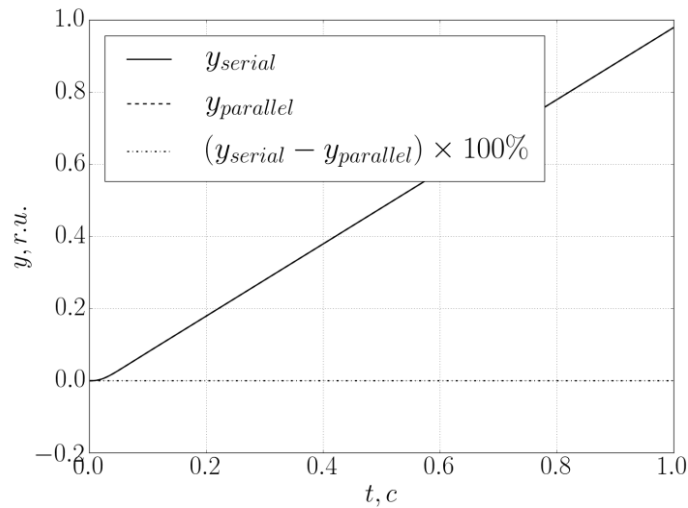

**Fig. 1.** Results of simulations serial and parallel models

Analysis of simulation's results makes it possible to conclude that serial and parallel models allow us to get the same results. Moreover, numerical solutions for both models coincide with great accuracy which is less than sample time.

### 3.3. A Generalized Algorithm for Transformation of Serial Models into Parallel Ones

Above-given theoretical and numerical results allow us to formulate an algorithm for transformation serial mathematical continuous time model into parallel discrete time one. There are two operations which can be concluded:

1) Operations which are performed in the continuous time domain, which has produced:

- State space equations for the considered object are written down.
- The transfer function is defined.
- Characteristic polynomial is analyzed and eigenvalues are defined.
- Partial fraction decomposition for transfer function is performed.
- State space equations for new virtual state variables are written down.

2) Operations which are performed in the discrete time domain, which is containing:

- Approximation formula for the derivative operator is chosen.
- State space equations are rewritten into the discrete time domain.
- Mesh equations are written down.

### 4. Conclusion

The mathematical model of the generalized linear dynamical system can be transformed into parallel form by performing partial fraction decomposition for its transfer function. Usage of this transformation allows us to simplify the order of considered dynamical system and study its dynamical and static characteristics by analyzing every parallel channel. One can get reducing of simulation time while the dynamic of every parallel channel is being calculated using multithreaded programming.

## References

[1] T. Sulman, O. Solovyova, and L. Katsnelson, "Combined mathematical model of the electrical and mechanical activity of the human cardiomyocyte," in *2018 Ural Symposium on Biomedical Engineering, Radioelectronics and Information Technology (USBEREIT)*, 2018, pp. 25–28, doi: https://doi.org/10.1109/USBEREIT.2018.8384541.

[2] R. Igual, C. Medrano, F. J. Arcega, and G. Mantescu, "Integral mathematical model of power quality disturbances," in *2018 18th International Conference on Harmonics and Quality of Power (ICHQP)*, 2018, pp. 1–6, doi: https://doi.org/10.1109/ICHQP.2018.8378902.

[3] F. Alhammadi, W. Waheed, B. El-Khasawneh, and A. Alazzam, "Mathematical model and verification of electric field and dielectrophoresis in a microfluidic device," in *2018 Advances in Science and Engineering Technology International Conferences (ASET)*, 2018, pp. 1–5, doi: https://doi.org/10.1109/ICASET.2018.8376900.

[4] M. Kasian and K. Kasian, "Diagnostic mathematical model of radio-electronic devices," in *2018 14th International Conference on Advanced Trends in Radioelecrtronics, Telecommunications and Computer Engineering (TCSET)*, 2018, pp. 766–770, doi: https://doi.org/10.1109/TCSET.2018.8336312.

[5] Y. Xu and C. Li, "Research on the Early Intelligent Warning System of Lost Circulation Based on Fuzzy Expert System," in *2018 International Conference on Intelligent Transportation, Big Data & Smart City (ICITBS)*, 2018, pp. 540–544, doi: https://doi.org/10.1109/ICITBS.2018.00142.

[6] S. Prykhodko, N. Prykhodko, L. Makarova, and K. Pugachenko, "Detecting outliers in multivariate non-Gaussian data on the basis of normalizing transformations," in *2017 IEEE First Ukraine Conference on Electrical and Computer Engineering (UKRCON)*, 2017, pp. 846–849, doi: https://doi.org/10.1109/UKRCON.2017.8100366.

[7] M. Awad, "New mathematical models to estimate wheat Leaf Chlorophyll Content based on Artificial Neural Network and remote sensing data," in *2016 IEEE International Multidisciplinary Conference on Engineering Technology (IMCET)*, 2016, pp. 86–91, doi: https://doi.org/10.1109/IMCET.2016.7777432.

[8] M. Zagirnyak, I. Zachepa, and V. Chenchevoi, "Estimation of induction generator overload capacity under connected direct current consumers," *Acta Tech.*, vol. 59, no. 2, pp. 149–169, 2014, available at : http://journal.it.cas.cz/59(14)2-Contents/59(14)2c.pdf.

[9] R. S. Voliansky and A. V. Sadovoi, "The transformation of linear dynamical object's equation to Brunovsky canonical form," in *2017 IEEE 4th International Conference Actual Problems of Unmanned Aerial Vehicles Developments (APUAVD)*, 2017, pp. 196–199, doi: https://doi.org/10.1109/APUAVD.2017.8308808.

[10] R. Voliansky, A. Sadovoi, and N. Volianska, "Interval model of the piezoelectric drive," in *2018 14th International Conference on Advanced Trends in Radioelecrtronics, Telecommunications and Computer Engineering (TCSET)*, 2018, pp. 1–6, doi: https://doi.org/10.1109/TCSET.2018.8336211.

[11] X. Zhi, W. Chunling, Q. Shuo, and M. Jin, "Research on mathematical model of electrode boiler based on neural network," in *2017 13th IEEE International Conference on Electronic Measurement & Instruments (ICEMI)*, 2017, pp. 281–285, doi: https://doi.org/10.1109/ICEMI.2017.8265792.

[12] Y.-L. Jin, A.-H. Tang, Y. Huang, X. Zheng, and Q.-S. Xu, "Research for Equivalent Mathematical Model of MMC-DPFC," in *2017 International Conference on Industrial Informatics - Computing Technology, Intelligent Technology, Industrial Information Integration (ICIICII)*, 2017, pp. 231–236, doi: https://doi.org/10.1109/ICIICII.2017.46.

[13] V. V. Mykhailenko, S. O. Buryan, T. B. Maslova, L. M. Naumchuk, J. M. Chunyak, and O. S. Charnjak, "Mathematical model of the semiconductor converter with twelve-zone regulation of output voltage," in *2017 IEEE First Ukraine Conference on Electrical and Computer Engineering (UKRCON)*, 2017, pp. 365–368, doi: https://doi.org/10.1109/UKRCON.2017.8100511.

[14] D. T. P. Wijesuriya, L. S. Wijesinghe, K. D. S. H. Wickramathilaka, D. M. Vithana, and H. Y. R. Perera, "A novel mathematical model to improve the power output of a solar panel using reflectors," in *2016*

*Electrical Engineering Conference (EECon)*, 2016, pp. 97–102, doi: https://doi.org/10.1109/EECon.2016.7830942.

[15] I. M. Buzurovic and S. Salinic, "A mathematical model of a novel automated medical device for needle insertions," in *2015 IEEE 15th International Conference on Bioinformatics and Bioengineering (BIBE)*, 2015, pp. 1–5, doi: https://doi.org/10.1109/BIBE.2015.7367697.

[16] X. An, B. Song, Z. Mao, and C. Ma, "The mathematical modeling of a novel anchor based on vortex induced vibration," in *OCEANS 2016 - Shanghai*, 2016, pp. 1–5, doi: https://doi.org/10.1109/OCEANSAP.2016.7485461.

[17] M. Antonova, E. Vasilieva, and I. Zhezhera, "The antisurge protection of a centrifugal compressor," in *2017 International Conference on Modern Electrical and Energy Systems (MEES)*, 2017, pp. 80–83, doi: https://doi.org/10.1109/MEES.2017.8248958.

[18] S. I. Gordon and B. Guilfoos, *Introduction to Modeling and Simulation with MATLAB®and Python*. Chapman and Hall/CRC, 2017, available at: https://www.taylorfrancis.com/books/9781498773881.

[19] N. Hara and K. Konishi, "On the use of intermediate solutions in parallel model predictive control based on matrix splitting," in *2017 11th Asian Control Conference (ASCC)*, 2017, pp. 2280–2285, doi: https://doi.org/10.1109/ASCC.2017.8287530.

[20] M. R. Rathomi and R. Pulungan, "A coarse-grained parallelization of genetic algorithms," *Int. J. Adv. Intell. Informatics*, vol. 4, no. 1, pp. 1–10, Apr. 2018, doi: https://doi.org/10.26555/ijain.v4i1.137.

[21] A. M. Yu, A. A. Gunkin, and A. S. Ananenkov, "Mathematical Modeling of Dynamics Parallel Manipulator-Tripod with Six Degrees of Freedom," *World Appl. Sci. J.*, vol. 30, no. 10, pp. 1298–1305, 2014, doi: https://doi.org/10.5829/idosi.wasj.2014.30.10.14170.

[22] G. Szederkényi, R. Lakner, and M. Gerzson, *Intelligent control systems: an introduction with examples*, vol. 60. Springer Science & Business Media, 2006, available at: https://books.google.com.

[23] D. Słota and R. Wituła, "Three Brick Method of the Partial Fraction Decomposition of Some Type of Rational Expression," 2005, pp. 659–662, doi: https://doi.org/10.1007/11428862_89.